\documentclass[12pt,a4paper]{article}

\usepackage{amsmath,amssymb}
\usepackage{mathptmx}
\usepackage[no-math]{fontspec}
\usepackage{mathspec}
\setmathrm[BoldFont=texgyretermes-bold.otf,
  ItalicFont=texgyretermes-italic.otf,
  BoldItalicFont=texgyretermes-bolditalic.otf]{texgyretermes-regular.otf}
\setmathsfont(Digits,Latin,Greek)[BoldFont=texgyretermes-bold.otf,
  ItalicFont=texgyretermes-italic.otf,
  BoldItalicFont=texgyretermes-bolditalic.otf]{texgyretermes-regular.otf}
\usepackage[a4paper,margin=2.5cm]{geometry}
\usepackage{graphicx,booktabs,array,tabularx,enumitem,titlesec,xcolor,caption}
\usepackage{tikz}
\usetikzlibrary{arrows.meta,positioning,calc,fit,backgrounds}
\usepackage[hidelinks]{hyperref}
\hypersetup{
  pdftitle={MatQuery: Execution-Grounded Natural-Language Querying over Heterogeneous Materials Databases},
  pdfauthor={Chen Jason Zhang and Qi Chen},
  pdfsubject={Technical report}
}
\usepackage{url}
\titleformat{\section}{\normalfont\large\bfseries}{\thesection}{0.8em}{\MakeUppercase}
\titleformat{\subsection}{\normalfont\normalsize\bfseries}{\thesubsection}{0.8em}{}
\titlespacing*{\section}{0pt}{12pt}{5pt}
\titlespacing*{\subsection}{0pt}{9pt}{3pt}
\setlist{nosep,leftmargin=1.5em}
\newcolumntype{Y}{>{\raggedright\arraybackslash}X}

\makeatletter
\newcommand{\refitem}[2]{\if@filesw\immediate\write\@auxout{\string\bibcite{#1}{#2}}\fi\textbf{[#2]}~}
\makeatother

\newcommand{\code}[1]{\texttt{#1}}

\begin{document}

\thispagestyle{plain}
\begin{center}
{\fontsize{19}{23.5}\selectfont\bfseries MatQuery: Execution-Grounded Natural-Language\\[2pt]
Querying over Heterogeneous Materials Databases\par}
\vspace{16pt}
{\large Chen Jason Zhang and Qi Chen\par}
\vspace{4pt}
{\itshape Department of Computing, The Hong Kong Polytechnic University\par}
{\itshape Hung Hom, Kowloon, Hong Kong\par}
\vspace{10pt}
{Technical Report\par}
\end{center}
\vspace{4pt}
\noindent\rule{\textwidth}{0.6pt}\par\vspace{2pt}
\noindent{\large\bfseries ABSTRACT}\par\vspace{4pt}
\noindent Materials scientists increasingly combine information from several public databases,
each with its own interface, schema, units and calculation conventions. MatQuery is a prototype and
experimental infrastructure for studying how a natural-language request can be turned into
queries over such databases and how the returned values can be interpreted. This report describes what has been built and the preliminary results obtained so far. The
prototype holds frozen,
self-verifying local copies of six public materials databases (about 10,000 records each), a
catalog of declared fields derived from them, a database-independent query representation with an
adapter for each database, and an eight-step pipeline in which a language model proposes and
deterministic program steps check, execute and assemble. A benchmark generator produces questions
whose reference queries are verified by execution on the frozen copies and whose wording is checked
by back-translation and review. An evaluation harness runs direct prompting, CHESS and Alpha-SQL on
the same frozen basis. Preliminary experiments on an 800-question development set show that
direct generation with three demonstrations reaches 82.4\% strict execution accuracy (56.5\%
without demonstrations), that a low-cost scorer selects the exact set of databases for 85.0\% of
complete requests with well-calibrated probabilities, and that band gaps differ by more than
0.5~eV in 31.6\% of composition-matched cross-database clusters. Two pilots show the limits of the
current approach: a probe-then-ask policy gave answers rated valid by both
reviewers in only 11 of 50 sessions, and
only 115 of 300 sampled record pairs could be compared by structure. The complete pipeline is in
smoke testing and has not yet been evaluated at scale.
\par\vspace{6pt}
\noindent{\bfseries KEYWORDS}\par\vspace{2pt}
\noindent Natural-language interfaces to databases; text-to-SQL; heterogeneous data integration;
query federation; materials databases; benchmark construction; data provenance.\par\vspace{2pt}
\noindent\rule{\textwidth}{0.6pt}

% =====================================================================================
\section{Introduction}
% =====================================================================================

A materials scientist who screens candidates for a new compound rarely finds everything in one
place. The Materials Project~\cite{mp}, OQMD~\cite{oqmd}, AFLOW~\cite{aflow}
and JARVIS-DFT~\cite{jarvis} publish properties computed by density-functional theory,
NOMAD~\cite{nomad} archives the raw output of many such calculations, and the Crystallography Open
Database (COD)~\cite{cod} collects crystal structures determined by experiment. Each database
answers its own kind of question well. A request such as ``which of these candidates have actually
been made, and how do their computed band gaps compare across databases?'' nevertheless requires
several of them at once, and the scientist must still find the right fields, align their units,
and decide whether two records describe the same material.

Natural-language interfaces to databases~\cite{nalir} promise to remove much of this work, and
agents such as LLaMP~\cite{llamp} and ChatMOF~\cite{chatmof} already ground a language model in
one materials source. Recent text-to-SQL
systems based on large language models perform well on benchmarks such as Spider~\cite{spider} and
BIRD~\cite{bird}, using decomposition~\cite{dinsql}, prompt design~\cite{dailsql}, schema
pruning~\cite{chess} or search~\cite{alphasql}. Those benchmarks, however, involve one relational
database at a time and a question whose meaning is complete. A materials request differs in three
ways. It may need several databases whose schemas share no column names. Its scientific meaning
depends on conventions, such as the calculation method behind a band gap, that the request often
leaves unstated. And a query that executes correctly does not establish that the values it returns
can be compared, because two records with the same formula may describe different crystal
structures.

MatQuery was built to study these problems on real data. Its design follows one principle: the
language model proposes, and the program checks. A model reads the request, chooses databases,
matches fields and writes queries. Deterministic steps verify each proposal against a catalog of
declared fields, run the query on a frozen local copy, recognize the same material across databases
and assemble an answer table that keeps every value's source. The same frozen copies serve a
benchmark generator and an evaluation harness, so that every system is compared on an identical
basis.

This report documents the prototype, the benchmark construction and the evaluation harness, and it
reports the preliminary results obtained so far. Section~\ref{sec:data}
describes the data, Section~\ref{sec:pipeline} the pipeline, Section~\ref{sec:bench} the benchmark
generator and Section~\ref{sec:eval} the evaluation harness. Section~\ref{sec:prelim} reports the preliminary results, Section~\ref{sec:status} the current status and the work
that follows, and
Section~\ref{sec:limits} the limits of the evidence.

\paragraph{Current status.}
The data layer, the field catalog, the query representation, the adapters, the eight-step pipeline,
the benchmark generator and the evaluation harness are implemented. The complete pipeline is in smoke testing on a
30-sample validation profile, and its full-scale runs and ablations are in progress. The preliminary
results in Section~\ref{sec:prelim} were obtained on a development set of 800 questions with direct
generation and with separate pilot configurations, not with the complete pipeline of
Section~\ref{sec:pipeline}. Independent expert review of the questions has not yet taken place, and the
pilot labels were produced by AI reviewers and then checked by team members with materials-science
backgrounds.

% =====================================================================================
\section{Data: Six Databases and Their Frozen Copies}
\label{sec:data}
% =====================================================================================

\subsection{The six sources}
The prototype draws on six public sources that differ in how they are accessed, how many fields
they declare and what kind of data they hold (Table~\ref{tab:sources}). Five hold computed data;
NOMAD is an archive of calculations from many groups and codes, and COD holds experimental
structures. Access ranges from an official Python SDK (Materials Project) and REST interfaces
(OQMD, NOMAD) to the AFLUX query language (AFLOW), a downloadable release (JARVIS-DFT) and a public
MySQL service (COD). The OPTIMADE specification~\cite{optimade,optimade2024} offers a common query
interface to several of these providers, but it standardizes how data are requested rather than
what the returned values mean, so MatQuery keeps a source-specific adapter for each database.

\begin{table}[t]
\centering
\caption{The six sources and their frozen copies in the current snapshot set. ``Fields'' counts the
fields each source's interface declares in the prototype's catalog.}
\label{tab:sources}
\begin{tabularx}{\textwidth}{@{}l Y r r Y@{}}
\toprule
Source & Native access & Records & Fields & Kind of data\\
\midrule
Materials Project~\cite{mp} & Python SDK & 10,000 & 33 & computed\\
OQMD~\cite{oqmd} & HTTP REST & 10,000 & 23 & computed\\
AFLOW~\cite{aflow} & AFLUX & 10,000 & 37 & computed\\
JARVIS-DFT~\cite{jarvis} & JSON release & 9,996 & 60 & computed\\
NOMAD~\cite{nomad} & JSON over HTTP & 10,000 & 28 & computed (archive)\\
COD~\cite{cod} & MySQL & 10,000 & 40 & experimental\\
\bottomrule
\end{tabularx}
\end{table}

\subsection{Frozen, self-verifying snapshots}
Every experiment runs on local copies rather than on the live services. A snapshot captures the
state of one upstream database at one time as a read-only copy that verifies itself: it stores the
normalized records, the raw payloads, a SQLite representation, summary statistics and a manifest
with file digests, and validation checks the record counts, the digests and the consistency of the
stored representations. Once a snapshot has been taken, all queries and evaluations read only from
it and make no network call. Live services change, and a result obtained today could not otherwise
be reproduced next month. The current snapshot set is identified as \code{materials-project-hard},
\code{oqmd-hard}, \code{aflow-hard}, \code{jarvis-hard}, \code{nomad-hard} and \code{cod-hard}. The
collection used for the preliminary results in Section~\ref{sec:prelim} holds 10,000 records per
source, selected by a deterministic diversity procedure with a fixed seed. Both collections are
development resources selected for coverage, not samples of the complete databases.

A normalized record carries a namespaced identifier, the database and snapshot identifiers, the
upstream record identifier, the formula and element set, structural descriptors, property values,
method metadata, the data origin, the source URL and a digest of the raw payload. Namespacing
prevents an identifier from one source from being confused with the same string in another.
Scientific values remain nullable: a source that does not report a quantity is not given a
substitute value, and in particular a missing band gap is never treated as zero. Method strings are
kept as the source wrote them. Normalization therefore gives the query layer a common interface,
while provenance keeps the path back to the original evidence.

\subsection{Two kinds of heterogeneity}
Asking one question across the six sources requires overcoming two differences. The first is
naming. The band gap is called \code{Egap} in AFLOW, \code{band\_gap} in the Materials Project and
OQMD, \code{optb88vdw\_bandgap} in JARVIS-DFT and a nested \code{\dots band\_gap.value} in NOMAD,
and COD does not declare it at all. A query must map each of these to the same physical quantity
before it can be written. The second is coverage. Of the six sources, all declare a space-group
number, five a band gap, four a formation energy, three an energy above the convex hull and only one
a dielectric constant. A quantity that a source does not declare is absent, not zero, and it must
not enter a comparison as if it were a value.

Units and quantities add a third layer. Formation energy per atom and energy above the hull are
both expressed in eV per atom, yet they answer different questions: the first compares a compound
with its elements, the second with the competing phases at the same composition, and both depend
on each database's reference states and corrections. A unit check alone therefore cannot establish
that two values may be compared.

\subsection{What real questions look like}
To see whether cross-database questions occur in practice, the team collected requests from
official documentation, community forums and the screening criteria of research papers. Three
examples show the range. The Materials Project documentation asks for ``band gaps for all materials
containing only Si and O''. A forum user asks how to obtain all available calculated band gaps of
MoS$_2$ with the $P6_3/mmc$ space group in the Materials Project. A screening study removes ``all
the entries from the dataset that have energy above hull $>$2 eV/atom''. The requests are short and
precise to their authors. They also show that real requirements are not always confined to one
database: a user who wants every available calculated value, or a structure confirmed by
experiment, needs evidence from several sources at once.

% =====================================================================================
\section{The MatQuery Pipeline}
\label{sec:pipeline}
% =====================================================================================

\subsection{Design principle}
\begin{figure}[t]
\centering
\begin{tikzpicture}[x=0.965cm,y=1cm,>=Latex,
  mstep/.style={draw,rounded corners=2pt,fill=black!10,align=center,text width=2.6cm,minimum height=1.6cm,inner sep=2pt},
  pstep/.style={draw,rounded corners=2pt,fill=white,align=center,text width=2.6cm,minimum height=1.6cm,inner sep=2pt},
  band/.style={draw,gray,rounded corners=3pt},
  blab/.style={rotate=90,font=\bfseries}]
% bands
\draw[band] (-0.1,0.05) rectangle (15.75,-2.45);
\draw[band] (-0.1,-3.0) rectangle (15.75,-6.45);
\draw[band] (-0.1,-7.05) rectangle (15.75,-9.3);
\node[blab] at (-0.45,-1.2) {Understand};
\node[blab] at (-0.45,-4.72) {Execute};
\node[blab] at (-0.45,-8.17) {Compose};
% understand
\node[pstep] (s1) at (1.45,-1.2) {1. Field table};
\node[mstep] (s2) at (4.75,-1.2) {2. Extract\\conditions};
\node[mstep] (s3) at (8.05,-1.2) {3. Link fields,\\one source\\at a time};
\node[mstep] (s4) at (11.35,-1.2) {4. Select\\sources};
\node[pstep,text width=1.9cm] (r) at (14.45,-1.2) {retrieval\\of stored\\values};
\draw[->] (s1)--(s2); \draw[->] (s2)--(s3); \draw[->] (s3)--(s4); \draw[->] (r)--(s4);
% understand -> execute
\draw[->] (s4.south) -- (s4.south |- 0,-2.72) -- (1.45,-2.72) -- (1.45,-3.85);
% execute
\node[mstep] (s5) at (1.45,-4.6) {5. Write one\\canonical query};
\node[pstep] (s5b) at (4.75,-4.6) {compile to\\each dialect};
\node[pstep] (s6) at (8.05,-4.6) {6. Execute on\\the frozen copy};
\node[mstep] (s6b) at (11.35,-4.6) {result feedback:\\answered?};
\draw[->] (s5)--(s5b); \draw[->] (s5b)--(s6); \draw[->] (s6)--(s6b);
\draw[->] (s6b.south) -- (s6b.south |- 0,-5.85) -- (s5b.south |- 0,-5.85) node[pos=0.5,below] {bounded revision or retry} -- (s5b.south);
% execute -> compose
\draw[->] (s6b.east) -- (13.6,-4.6) -- (13.6,-6.75) -- (3.2,-6.75) -- (3.2,-7.55);
% compose
\node[mstep,text width=5.9cm,minimum height=1.45cm] (s7) at (3.2,-8.2) {7. Choose the merge operator:\\union, intersection, difference\\or comparison};
\node[pstep,text width=7.2cm,minimum height=1.45cm] (s8) at (11.5,-8.2) {8. Reconcile identity; compose the\\answer table with source, method\\and record identifier};
\draw[->] (s7)--(s8);
% legend
\node[draw,fill=black!10,minimum width=0.55cm,minimum height=0.35cm] at (4.0,-9.8) {};
\node[anchor=west] at (4.35,-9.8) {model call};
\node[draw,fill=white,minimum width=0.55cm,minimum height=0.35cm] at (8.0,-9.8) {};
\node[anchor=west] at (8.35,-9.8) {program step};
\end{tikzpicture}
\caption{The eight steps of the current MatQuery pipeline. Shaded boxes are model calls; white boxes
are program steps that check, execute or assemble.}
\label{fig:pipeline}
\end{figure}
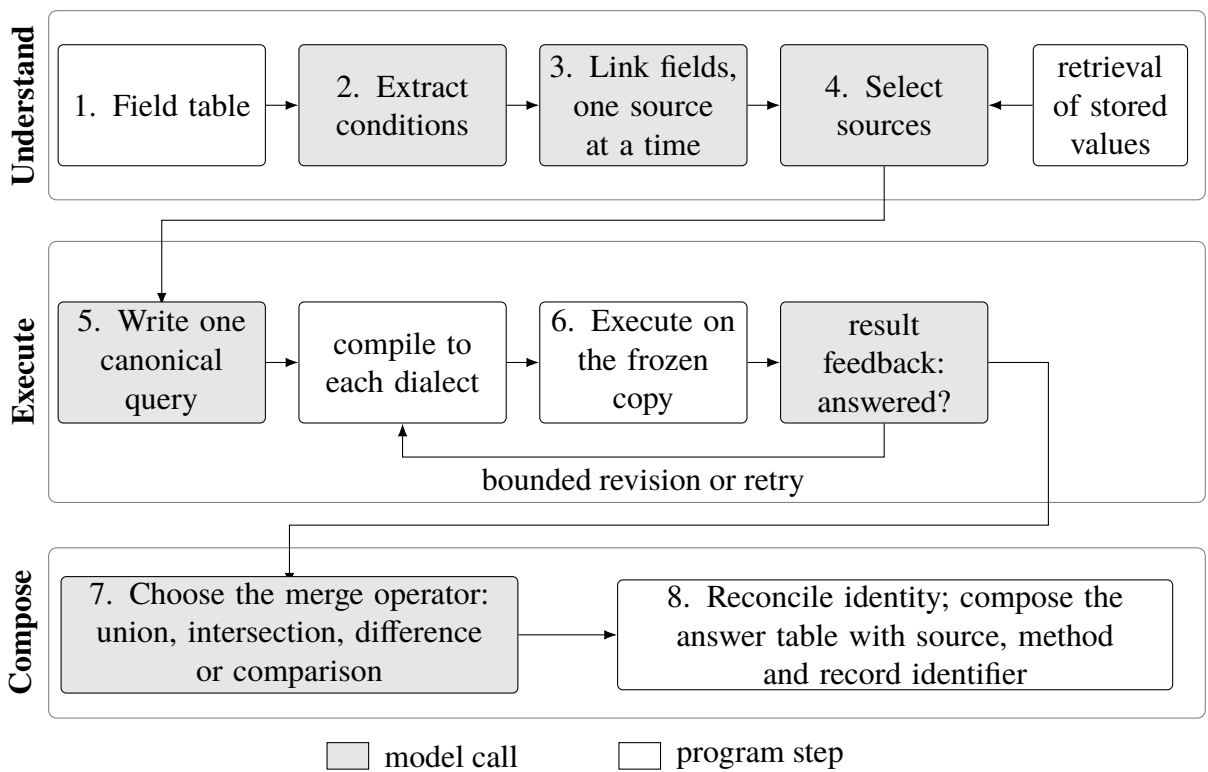

MatQuery turns a materials question into queries that run on several databases and returns a
result table in which every value carries its source. The work is divided between two kinds of
step (Figure~\ref{fig:pipeline}). Model calls interpret the question, choose databases, match
fields, write the query and judge the result. Program steps check that each proposal is legal,
execute queries on the frozen copies, recognize the same material across databases and assemble
the answer. A model call may be wrong, so its output is never executed or merged without a program
step that checks it first.

\subsection{Understanding the request}
\emph{Step 1: field table.} A program merges the fields declared by the six databases into one
catalog. For each field it records the meaning, the unit and physical dimension, a plausible
physical range, whether a calculation method must be stated with it, which databases hold it and
which of them allow it to be used as a filter. The catalog is built once per snapshot set and is
the reference against which every later proposal is checked.

\emph{Step 2: extracting the conditions.} One model call reads the question together with the
catalog and produces a structured query specification: the target, the property conditions with
their operators, values and units, composition and structure constraints, the requested operation
and any sources the user named. The specification uses the intermediate representation described
in Section~\ref{sec:ir}, so that what the user asked for is separated from how any one database
expresses it. A retrieval step gathers supporting evidence from the actual columns and stored
values, such as formulas, element sets and space groups that occur in the data.

\emph{Step 3: linking fields one source at a time.} For each candidate database, a separate model
call receives only that database's fields and selects those the specification needs. Presenting the fields of all six sources at once dilutes
the model's attention and invites a column name from one database to be used in another. Linking
one source at a time keeps each choice clean and lets each source report its fields under its own
names.

\emph{Step 4: selecting the sources.} A model call proposes which databases are needed and gives a
reason for each. A program step then checks coverage: every selected database must hold the fields
that its part of the request uses, and every required property must be covered by at least one
selected database. A selection that fails the check is rejected, not executed.

\subsection{Executing the query}
\emph{Step 5: one canonical query, compiled to each dialect.} The model writes a single query in a
database-independent form. A compiler translates it step by step into each target database's
executable statement on the local copy. When compilation fails, for example because of a syntax
error or a field that does not exist in that database, the specific error is returned to the model,
which may revise the query a bounded number of times. Revision can be switched off for ablation.

\emph{Step 6: execution with feedback.} The compiled query runs on the frozen copy. The program
returns the execution status, the number of rows, the first rows and the result columns to the
model, which judges whether the query has actually answered the question. If it has not, the model
may revise the query within the same bounded budget. Execution feedback can also be switched off.

\subsection{Composing the answer}
For a request that spans several databases, each database first returns its own part of the result.
Writing $q_s$ for the query branch sent to source $s$, $D_s$ for its frozen copy and $\rho$ for the
matching policy, the prototype computes
\begin{equation}
R_s=q_s(D_s)\ \ (s\in S),\qquad
\mathcal{C}=\operatorname{reconcile}\Big(\bigcup_{s\in S}R_s;\ \rho\Big),\qquad
Y=\operatorname{op}(\mathcal{C},S),
\label{eq:federation}
\end{equation}
where $\mathcal{C}$ is a set of groups of records judged to describe the same material and
$\operatorname{op}$ is the requested operation. In \emph{Step 7} a model call chooses
$\operatorname{op}$: a union keeps the groups contributed by any branch, an intersection keeps
groups present in every required source, a difference keeps groups present only in the first, and
a comparison keeps the grouped records side by side. In \emph{Step 8} a program performs the
reconciliation and builds the final table. The plan is executed as a directed acyclic graph of query
nodes, a reconciliation node and an output node. Equation~\eqref{eq:federation} describes record
integration at the application layer; it is not a relational join across remote databases.

Reconciliation uses the evidence the records carry: normalized formulas, element sets, space
groups, lattice parameters and, where available, structure fingerprints. In composition mode,
records are grouped when formula and element set agree. In structural mode, structural agreement is
also required. The rule that produced each group is kept with the result, so that a grouping by
composition can always be told apart from one supported by structure. The final table keeps the
database name, upstream record identifier, snapshot, method string, data origin, source URL and
value for every entry. Values from different databases are shown side by side and are never
averaged automatically, and recorded issues concerning units, missing method information or
reconciliation are attached for the reader to inspect.

\subsection{The intermediate representation}
\label{sec:ir}
The query representation separates the scientific request from the syntax of any one source
(Table~\ref{tab:ir}). A property constraint records the property, operator, value or range, unit,
method and conditions. Composition and structure constraints describe the material, and
projection, ordering, aggregation and route information describe how the answer is assembled.
Deterministic validation resolves property aliases against the catalog, converts supported units
to canonical units and identifies conditions that must be stated. The representation can also
express a clarification requirement and accept a structured answer, which is the interface needed
for experiments in which the system asks the user to settle part of a request.

\begin{table}[t]
\centering
\caption{Principal components of the intermediate representation.}
\label{tab:ir}
\begin{tabularx}{\textwidth}{@{}>{\raggedright\arraybackslash}p{3.3cm} Y Y@{}}
\toprule
Component & Information represented & Example\\
\midrule
Target and intent & material, structure or calculation record; filter, rank or aggregate & summarize band gaps by crystal system\\
Property predicates & property, operator, value or bounds, unit, method, conditions & band gap between 1.1 and 1.7 eV\\
Composition & formula, required elements, excluded elements & halides without Pb\\
Structure & crystal system, space group, lattice lengths, prototype & space group $P6_3/mmc$\\
Answer operation & projection, sort, limit, aggregate functions, grouping & top seven records by band gap\\
Route and provenance & database, snapshot, native representation & bind a branch to \code{jarvis-hard}\\
\bottomrule
\end{tabularx}
\end{table}

% =====================================================================================
\section{Benchmark Construction}
\label{sec:bench}
% =====================================================================================

\subsection{Approach}
No public benchmark poses natural-language questions over several materials databases with
verified answers, so MatQuery generates its own. The approach reverses the usual order. Instead of
writing questions and then searching for their answers, the generator first fixes a reference query
that executes correctly on the frozen copies, and only then asks a model to write the question a
researcher would ask. The question is accepted only if it can be read back into the same query and
passes a review of fidelity and naturalness. Every accepted sample therefore has a reference answer
that has been computed, not merely asserted.

\subsection{Six stages}
\emph{Capability pre-check.} For each coverage slot, a combination of databases, properties and
operations, the generator first confirms that the required fields exist and can be filtered in the
relevant databases. For slots that span databases, it also confirms that the databases share
materials that can be matched. A slot that fails stops here, and a capability report records the
outcome.

\emph{Reference pre-check.} The generator then collects enough candidate reference queries for each
slot, each of which must actually run. It draws a few more candidates than needed and filters them
against a set of exclusion fingerprints, so that a query meaning already used elsewhere is not
produced again.

\emph{Reference synthesis.} Each candidate is executed on the snapshots. A candidate is discarded if
execution fails or if its result does not meet the slot's coverage requirement, for example a
comparison that returns records from only one database.

\emph{Verbalization.} A model receives a reader-facing description of the reference query, its
research goal and its scientific conditions, with property names, comparison, values and units in
natural language, together with a style instruction. It writes one question per style. The current
generator offers seven styles: explicit expert, natural scholarly, concise, colloquial, incomplete,
ambiguous and multilingual. The incomplete and ambiguous styles deliberately leave out or blur part
of the request and provide material for clarification experiments.

\emph{Back-translation.} A separate model call reads the question back into a structured
specification, and a program compares it field by field with the reference. Any mismatch rejects
the question.

\emph{Alignment review.} For questions that pass the field comparison, another model call acts as a
reviewer. It judges whether the meaning is preserved, whether the request is scientifically sound,
and whether the parts meant to be vague are actually vague. It also gives a confidence and a
naturalness score, and thresholds on these decide acceptance.

\subsection{Two generations of the benchmark}
The development set used for the preliminary results in Section~\ref{sec:prelim} contains 800 complete base questions,
20 in each of 40 coverage slots, of which 460 address one database and 340 several. It also contains
120 incomplete variants linked to their base questions, giving 920 question versions in all. Of
these, 447 were produced by a controlled template renderer and 473 by model-based verbalization.
Table~\ref{tab:taskforms} shows the operations it covers.

\begin{table}[t]
\centering
\caption{Task forms in the 800-question development set, recomputed from the saved
reference plans.}
\label{tab:taskforms}
\begin{tabularx}{\textwidth}{@{}Y r r Y@{}}
\toprule
Task form & Questions & Share & Main requirement\\
\midrule
Single-source filtering & 340 & 42.5\% & composition, structure and property predicates\\
Single-source ranking & 40 & 5.0\% & ordering and a result limit\\
Single-source aggregation & 80 & 10.0\% & summaries and groups\\
Multi-source comparison & 180 & 22.5\% & source-specific records side by side\\
Multi-source union & 100 & 12.5\% & combine contributions of branches\\
Multi-source intersection & 40 & 5.0\% & groups present in all required branches\\
Multi-source difference & 20 & 2.5\% & ordered difference between branches\\
\midrule
Total & 800 & 100\% & 40 coverage slots\\
\bottomrule
\end{tabularx}
\end{table}

The current generator implements the six stages above. Its smoke-test profile,
\code{validation\_\allowbreak small}, is summarized in Table~\ref{tab:smoke}. It contains 30 samples, one per
coverage slot without repetition, of which 25 address one database and 5 several; 28 are marked
hard and 2 easy. The 30 samples have 196 question variants across the seven styles. The reference
answers draw on the Materials Project for 11 samples, on AFLOW, JARVIS-DFT, NOMAD and COD for 4
each, and on OQMD for 3. The profile is marked eligible for release. It exists to verify the
pipeline from end to end, not to estimate accuracy.

\begin{table}[t]
\centering
\caption{The current smoke-test profile \code{validation\_\allowbreak small}.}
\label{tab:smoke}
\begin{tabularx}{\textwidth}{@{}>{\raggedright\arraybackslash}p{5.6cm} Y@{}}
\toprule
Statistic & Value\\
\midrule
Samples & 30\\
Coverage slots & 30 (one sample per slot, no repetition)\\
Single-source / multi-source & 25 / 5\\
Difficulty (hard / easy) & 28 / 2\\
Question variants & 196\\
Styles per sample & 7 (expert explicit, scholarly natural, concise, colloquial, incomplete, ambiguous, multilingual)\\
Databases in reference answers & Materials Project 11; AFLOW 4; JARVIS-DFT 4; NOMAD 4; COD 4; OQMD 3\\
Release status & eligible for release\\
\bottomrule
\end{tabularx}
\end{table}

A separate pool of demonstrations supports the few-shot baselines. Demonstrations come from an
independently released, reviewed pool; each is routed to the profile of its own database, and a
leakage check refuses any candidate that shares a sample identifier, a reference query or an unseen
intent with the evaluation set. The sample under evaluation is always excluded from its own
candidate set.

% =====================================================================================
\section{Evaluation Harness}
\label{sec:eval}
% =====================================================================================

\subsection{One frozen basis for every system}
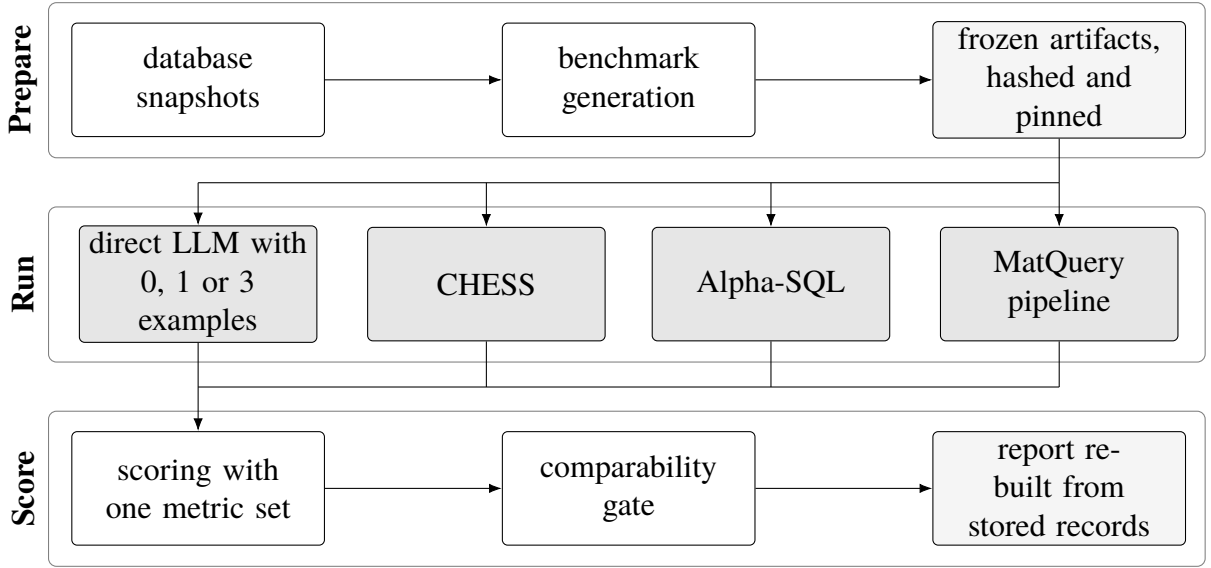
\begin{figure}[t]
\centering
\begin{tikzpicture}[x=0.965cm,y=1cm,>=Latex,
  box/.style={draw,rounded corners=2pt,align=center,text width=3.2cm,minimum height=1.5cm,inner sep=2pt},
  sbox/.style={draw,rounded corners=2pt,align=center,text width=3.0cm,minimum height=1.5cm,fill=black!10,inner sep=2pt},
  band/.style={draw,gray,rounded corners=3pt},
  blab/.style={rotate=90,align=center,font=\bfseries}]
\draw[band] (-0.1,0.1) rectangle (15.75,-1.95);
\draw[band] (-0.1,-2.6) rectangle (15.75,-4.65);
\draw[band] (-0.1,-5.3) rectangle (15.75,-7.35);
\node[blab] at (-0.45,-0.92) {Prepare};
\node[blab] at (-0.45,-3.62) {Run};
\node[blab] at (-0.45,-6.32) {Score};
\node[box] (a1) at (1.95,-0.92) {database\\snapshots};
\node[box] (a2) at (7.85,-0.92) {benchmark\\generation};
\node[box,fill=black!4] (a3) at (13.75,-0.92) {frozen artifacts,\\hashed and pinned};
\draw[->] (a1)--(a2); \draw[->] (a2)--(a3);
\node[sbox] (b1) at (1.95,-3.62) {direct LLM with\\0, 1 or 3 examples};
\node[sbox] (b2) at (5.9,-3.62) {CHESS};
\node[sbox] (b3) at (9.8,-3.62) {Alpha-SQL};
\node[sbox] (b4) at (13.75,-3.62) {MatQuery\\pipeline};
\draw (a3.south) -- (13.75,-2.28);
\draw (1.95,-2.28) -- (13.75,-2.28);
\foreach \b in {b1,b2,b3,b4} {\draw[->] (\b.north |- 0,-2.28) -- (\b.north);}
\node[box] (c1) at (1.95,-6.32) {scoring with\\one metric set};
\node[box] (c2) at (7.85,-6.32) {comparability\\gate};
\node[box,fill=black!4] (c3) at (13.75,-6.32) {report rebuilt from\\stored records};
\draw[->] (c1)--(c2); \draw[->] (c2)--(c3);
\foreach \b in {b1,b2,b3,b4} {\draw (\b.south) -- (\b.south |- 0,-4.98);}
\draw (1.95,-4.98) -- (13.75,-4.98);
\draw[->] (1.95,-4.98) -- (c1.north);
\end{tikzpicture}
\caption{The evaluation harness. Shaded boxes call a language model; all systems read the same
pinned snapshots and questions.}
\label{fig:harness}
\end{figure}

The harness has three stages (Figure~\ref{fig:harness}). It first prepares the evidence: the
snapshots and the generated questions are hashed and pinned, and every run records which versions
it used. It then runs each system on that identical basis. Finally it scores every prediction
against the frozen reference with the same set of metrics and renders the tables from the stored
records. A comparability gate allows two runs to be compared only when their basis is identical, so
a difference in a table cannot come from a change in the data or the questions.

\subsection{Systems compared}
\emph{Direct prompting} is the simplest baseline. Without examples, the model reads the published
capability catalog of the six databases, chooses databases and writes one SQLite query, which the
program executes on the frozen copies while keeping an auditable record. The one-shot and few-shot
variants add exactly one, or by default three, demonstrations from the separate pool and otherwise
follow the identical chain of routing, generation, execution and storage. The few-shot variant may
write a separate branch query for each routed database, with one candidate and no revision per
branch.

\emph{CHESS}~\cite{chess} decomposes text-to-SQL into information retrieval, schema selection,
candidate generation and natural-language unit tests. The harness runs the authors' official
implementation: it prepares input in the format CHESS expects, applies compatibility patches in a
disposable copy, lets CHESS solve its questions with the same language model, and replays the
returned queries on the frozen copies for scoring.

\emph{Alpha-SQL}~\cite{alphasql} searches over reasoning actions with Monte Carlo tree search and
execution rewards. The harness runs the published source unchanged: it exports the questions as a
BIRD-compatible batch, runs the upstream scripts in an isolated runtime, imports the predicted
queries and replays them on the frozen copies. Its search makes it the slowest system in the
harness.

Both external systems were designed for one relational database at a time. Running them on the
same basis shows how far a strong single-database method goes on cross-database materials
questions, and it ensures that MatQuery is compared with the strongest available alternatives
rather than only with its own variants.

\subsection{Metrics}
The primary metric is execution accuracy. For a sample set $\mathcal{S}$, system prediction
$\hat y_i$, reference query $y_i$ and database $D_i$,
\begin{equation}
\mathrm{EX}=\frac{1}{|\mathcal{S}|}\sum_{i\in\mathcal{S}}
\mathbf{1}\big[\operatorname{Exec}(\hat y_i,D_i)\equiv\operatorname{Exec}(y_i,D_i)\big],
\label{eq:ex}
\end{equation}
where $\equiv$ denotes equivalence of result sets under the task's comparison rule: ordered record
identifiers for ranking, matching hashes and columns for aggregates, and record-identifier sets
otherwise. The second metric is routing accuracy, the share of samples in a set $\mathcal{R}$ for
which the selected set of databases equals the reference set,
\begin{equation}
\mathrm{Route}=\frac{1}{|\mathcal{R}|}\sum_{i\in\mathcal{R}}
\mathbf{1}\big[\hat{\mathcal{D}}_i=\mathcal{D}_i^{\mathrm{gold}}\big].
\label{eq:route}
\end{equation}
Exact agreement of the whole set matters because omitting one database can change a cross-database
answer even when most individual choices are right. Each prediction record also stores the query,
the returned rows, the tokens used and the cost, so that accuracy can be reported against cost.

Neither metric measures scientific validity. A result set equal to the reference shows agreement
under the saved scoring rule. It does not show that the methods behind the values are compatible
or that grouped records share a crystal structure. Those questions are examined separately
(Section~\ref{sec:audit}).

\subsection{Ablation switches}
The pipeline exposes switches that turn individual mechanisms on and off, so that their effect can
be measured on the same basis (Table~\ref{tab:switches}).

\begin{table}[t]
\centering
\caption{Ablation switches of the current pipeline.}
\label{tab:switches}
\begin{tabularx}{\textwidth}{@{}>{\raggedright\arraybackslash}p{5.6cm} r Y@{}}
\toprule
Switch & Default & Effect\\
\midrule
\code{MATQUERY\_\allowbreak SQL\_\allowbreak REVISION\_\allowbreak LIMIT} & 2 & maximum revisions after a compilation failure; 0 gives a single attempt\\
\code{MATQUERY\_\allowbreak RESULT\_\allowbreak FEEDBACK\_\allowbreak LIMIT} & 2 & maximum rounds of execution feedback; 0 disables it\\
\code{MATQUERY\_\allowbreak RESULT\_\allowbreak FEEDBACK\_\allowbreak SAMPLE\_\allowbreak ROWS} & 5 & result rows shown to the model\\
\code{MATQUERY\_\allowbreak SCHEMA\_\allowbreak LINK\_\allowbreak PER\_\allowbreak DATABASE} & true & link fields one database at a time; false links all six together\\
\code{MATQUERY\_\allowbreak IDENTITY\_\allowbreak RECONCILIATION} & true & group the same material across databases; false disables grouping\\
\bottomrule
\end{tabularx}
\end{table}

\subsection{Computing environment}
All current experiments run on a single laptop: an ASUS TX Air FA401KM with Windows~11 (build
26200), an AMD Ryzen AI~7 H~350 processor (8 cores, 16 threads), 32~GiB of memory, an NVIDIA GeForce
RTX~5060 Laptop GPU with 8~GiB of memory (driver 577.05, CUDA~12.9), a 1~TB NVMe drive, CPython
3.12.13 and \code{uv} 0.12.21. The dialogue model runs on a remote service; the laptop performs
orchestration, query execution and local embedding inference, and the resident embedding model
occupies 7,579~MiB of GPU memory. A complete harness, including six database copies, therefore fits on one laptop, although no part
of the current pipeline yet runs its dialogue model locally.

% =====================================================================================
\section{Preliminary Results}
\label{sec:prelim}
% =====================================================================================

This section reports the preliminary results. They were obtained on the 800-question development
set and the snapshot collection described in Sections~\ref{sec:data} and~\ref{sec:bench}, using
direct generation and separate pilot configurations rather than the complete pipeline of
Section~\ref{sec:pipeline}. The questions have not yet been reviewed by independent experts, so the
results describe behavior on this development resource.

\subsection{Direct query generation on 800 questions}
The main runs used a remote model through the endpoint alias \code{deepseek-flash}, with either no
demonstrations or three. Table~\ref{tab:full800} gives the results. With three demonstrations,
strict execution accuracy, which requires successful execution and a matching result hash, rose
from 452 to 659 of 800 questions. The exact source set was chosen correctly in about 90\% of
questions in both runs. Of the 258 questions that became correct only with demonstrations, 216 had
the correct source set in both runs, which places most of the gain in query construction rather
than in source choice.

\begin{table}[t]
\centering
\caption{Direct-generation results on the same 800 complete questions (denominators 800).}
\label{tab:full800}
\begin{tabularx}{\textwidth}{@{}Y r r@{}}
\toprule
Measure & No demonstrations & Three demonstrations\\
\midrule
Execution success & 515 (64.4\%) & 675 (84.4\%)\\
Strict execution accuracy & 452 (56.5\%) & 659 (82.4\%)\\
Result equivalence & 459 (57.4\%) & 666 (83.3\%)\\
Exact source-set match & 717 (89.6\%) & 718 (89.8\%)\\
\midrule
Strict accuracy, single-source (of 460) & 262 (57.0\%) & 387 (84.1\%)\\
Strict accuracy, multi-source (of 340) & 190 (55.9\%) & 272 (80.0\%)\\
\bottomrule
\end{tabularx}
\end{table}

The gain was uneven across operations. Aggregation rose from 20 to 75 of 80 questions and
cross-source comparison from 83 to 144 of 180, while ranking did not improve (20 and 19 of 40).
Demonstrations and evaluation questions share task families, so these figures describe transfer
within the development set and not to new kinds of request.

\subsection{Inexpensive source scoring}
A separate pilot posed one yes-or-no question per database for each of the 920 question versions,
5,520 decisions in all, and read the answer probability without generating text. Exact agreement
with the reference source set was 680 of 800 on the complete questions (85.0\%) and 800 of 920
across all versions (87.0\%). The area under the ROC curve for single decisions was 0.995, and
after temperature scaling~\cite{guo2017} fitted in two folds grouped by base question, the expected
calibration error was 0.0125 (0.0423 before scaling). The recorded provider cost of all 5,520
decisions was US\$0.032; six decisions shared one request with a median duration of 383~ms, and a
decision used on average 139 input tokens. The pilot shows that such a decision can be made cheaply
and with well-calibrated probabilities. On complete questions, however, its exact-set agreement was
below that of the generative pipeline (85.0\% against about 90\%), so its case rests on cost and
calibration, not on accuracy.

\subsection{Cross-source audit}
\label{sec:audit}
A deterministic audit of the six snapshots, independent of any language model, measured how much
values disagree for records grouped by composition. Formula normalization produced 5,732 clusters
containing records from at least two sources, each with at most one record per source. In 1,036 of
the 3,275 clusters with at least two band-gap values (31.6\%), the band gaps differed by more than
0.5~eV. The share was 175 of 740 (23.6\%) when only records reporting the same space group were
compared. Formation energies differed by more than 0.1~eV per atom in 55.7\% of eligible clusters,
and the stored hull-related values by more than 0.05~eV per atom in 79.0\%. Disagreement of this kind has been documented
for databases that use similar methods~\cite{hegde2023}. Every cluster contained more than one
recorded method string, but the strings mean different things: OQMD records labels
such as \code{static}, NOMAD records program names such as VASP, and JARVIS-DFT records the
functional. String inequality is therefore heterogeneity in metadata, not proof of incompatible
methods.

A single traced case shows why identity matters. One SiO$_2$ cluster contains records from AFLOW,
JARVIS-DFT, the Materials Project and OQMD with band gaps of 0.000, 5.469, 5.516 and 4.606~eV. The
three records that store a numerical space group report three different ones (70, 141 and 58). The
zero is present in the AFLOW payload itself and is not a substituted missing value. The four
records share a composition but not, on the evidence stored, a structure, and an interface that
returned one unqualified number would hide this.

\subsection{Fifty-session clarification pilot}
Before any model call, 50 sessions were fixed: 20 complete requests and 30 linked incomplete
variants. Executing candidate values on the snapshots labelled 15 omitted slots as changing the
result, 20 as not changing it, and 10 as already settled by the text. Six arms were compared on the
same sessions (Table~\ref{tab:session}). Validity was labelled by two separately blinded AI
reviewers and then checked by team members with materials-science backgrounds; independent expert
adjudication has not taken place.

\begin{table}[t]
\centering
\caption{The 50-session pilot. ``Valid'' gives the counts of the two AI
reviewers; ``invalid'' counts answers flagged by either. Denominators are 50 except for
questions.}
\label{tab:session}
\begin{tabularx}{\textwidth}{@{}Y r r r r r@{}}
\toprule
Arm & Executed & Ref. equiv. & Valid & Invalid & Questions\\
\midrule
A: direct, no questions & 42 & 37 & 19 / 19 & 20 & 0\\
B0: native framework, with context & 7 & 7 & 7 / 5 & 2 & 0\\
B: adapted, no questions & 2 & 2 & 2 / 2 & 0 & 0\\
C: ask & 16 & 15 & 15 / 14 & 1 & 77\\
D: probe, then ask & 14 & 13 & 11 / 13 & 3 & 69\\
E: direct agent with ask tool & 38 & 38 & 29 / 21 & 17 & 18\\
\bottomrule
\end{tabularx}
\end{table}

The pilot did not support the probe-and-ask design in its tested form. Arm D produced 11 answers
rated valid by both reviewers, with an exact 95\% interval of 11.5--36.0\%, and failed its
pre-specified gate of at least 25 valid and at most 2 invalid answers. Most of its failures (35 of
39) were failures to generate, compile or execute a plan, which point to the shared three-example
adaptation: it executed in 2 of 50 sessions where direct generation executed in 42. The questions
were also poorly targeted: 65 of the 77 questions asked by arm C and 57 of the 69 asked by arm D did
not concern an omitted slot that changed the result. The direct arms, by contrast, returned an
invalid answer in 17 to 20 of 50 sessions. Arm E is descriptive only, because the scripted replies
missed or under-answered ten of its questions. The two reviewers agreed on 273 of 300 labels
(Cohen's $\kappa=0.836$).

\subsection{Three-hundred-pair structure audit}
A sample of 300 record pairs was frozen: 150 pairs linked by the composition matcher, 100 unlinked
pairs with the same reported space group and 50 unlinked pairs with different space groups. Only
JARVIS-DFT kept full atomic coordinates in the frozen payloads, so on the snapshots alone every pair
was ``cannot tell''. Retrieving public structures by the saved record identifiers supplied both
structures for 115 pairs. Automatic labels from the pymatgen structure matcher~\cite{pymatgen} are
given in Table~\ref{tab:struct}. Among the linked pairs that could be resolved, 13 were the same
structure and 26 different.

\begin{table}[t]
\centering
\caption{Automatic structure labels for the 300 sampled pairs (pymatgen \code{StructureMatcher}
with volume scaling, after supplementary retrieval). The labels are a review aid, not ground
truth.}
\label{tab:struct}
\begin{tabularx}{\textwidth}{@{}Y r r r r@{}}
\toprule
Stratum & Pairs & Same & Different & Cannot tell\\
\midrule
Linked by composition & 150 & 13 & 26 & 111\\
Unlinked, same space group & 100 & 35 & 17 & 48\\
Unlinked, different space group & 50 & 0 & 24 & 26\\
\midrule
Total & 300 & 48 & 67 & 185\\
\bottomrule
\end{tabularx}
\end{table}

Two AI reviews of 78 selected pairs agreed on 71 ($\kappa=0.849$), and the labels were checked by
team members with materials-science backgrounds. The resulting estimates of linkage quality are
uncertain. The design-weighted precision of composition links is 33.1\% (95\% interval 19.6--47.8\%)
on the resolved pairs, but because structures were missing in a non-random way the
partial-identification range is 8.5--82.8\%. Only two pairs judged to be the same structure had two
band-gap values, so the audit cannot yet say how large the disagreement is between records known
to describe the same crystal.

\subsection{What the preliminary results establish}
Three conclusions survive the limits of these data. First, a substantial share of cross-database
requests can be turned into queries that return the reference result, but this establishes query
correctness and not scientific comparability. Second, values grouped by composition disagree often
enough, and composition links are unreliable enough, that identity must be judged by structure
before values are compared. Third, the simple probe-and-ask design failed in its tested form, mainly
because the adapted generation path broke and the questions asked were poorly targeted, not because
asking is unnecessary: the arms that asked nothing returned invalid answers in up to 40\% of
sessions. These findings shaped the pipeline of Section~\ref{sec:pipeline}, which links fields one database at a time,
checks coverage before execution, and revises queries with bounded feedback.

% =====================================================================================
\section{Current Status and Next Steps}
\label{sec:status}
% =====================================================================================

\subsection{Where the work stands}
The complete pipeline, the benchmark generator and the harness with all baselines are
implemented and in smoke testing on the \code{validation\_\allowbreak small} profile. Three steps
remain before results can be reported: completing the smoke-test validation, running the
large-sample experiments together with the ablations of Table~\ref{tab:switches}, and evaluating
and assembling all results. Alpha-SQL is the slowest system to run because of its tree search. When these runs are complete, a revision of this report will give their results alongside the
preliminary results of Section~\ref{sec:prelim}.

\subsection{Open items}
Several items must be settled before the benchmark can support confirmatory claims. The questions
have passed automatic checks and model-based review but no independent expert review, which is the first step planned with materials scientists. Structures must be re-acquired for all six sources, since
only JARVIS-DFT kept coordinates in the snapshots used for the preliminary results. The demonstration pool and the
evaluation set share task families, so a held-out split by family is needed to measure transfer.
Finally, the preliminary results were not produced by the complete pipeline, so they cannot be
read as its performance.

\subsection{Research directions supported by the prototype}
The components described in this report support four lines of further work. First, the field
catalog and the execution of candidate values on the snapshots make it possible to decide, from
the data, whether an unstated part of a request actually changes the answer, and therefore whether
it is worth asking the user about it. Second, the per-prediction records of tokens, model calls,
executions and cost make it possible to study how a fixed budget of computation and user
interaction should be spent within a session. Third, the typed steps of the pipeline (condition
extraction, field linking, source selection and the choice of merge operator), together with traces
verified by execution, form narrow decisions and training data on which smaller, locally deployed
models can be adapted and measured. Fourth, the compiler, bounded revision and execution feedback
are the starting point of a local runtime in which deterministic checks compensate for a weaker
model. The preliminary pilots set out what such work must improve on: clarification questions that
target the parts of a request that change the result, and a generation path that does not break
when it is adapted.

% =====================================================================================
\section{Limits of the Evidence and Reproducibility}
\label{sec:limits}
% =====================================================================================

The evidence in this report has a defined scope. The data are selected frozen subsets of six
sources, not samples of the complete databases. The questions are generated and checked
automatically, and independent expert review is pending. The pilot labels were produced by AI
reviewers and checked by team members with materials-science backgrounds, which is an internal
review rather than independent adjudication. Supplementary structures were retrieved after the
original snapshot freeze. Demonstrations and evaluation questions share task families. The model
endpoint was named by an alias that does not identify an immutable model revision, so the saved
outputs, not the alias, are the definitive record of each run. All figures therefore describe
behavior on these resources and are not general accuracy estimates for scientific queries.

Every reported figure is traceable to stored artifacts: snapshot manifests with file digests,
the benchmark manifest with its samples and reference plans, per-question prediction records with
queries, rows, tokens and cost, decision scores, and the deterministic audit report. The harness
rebuilds its tables from these records, and the comparability gate prevents results obtained on
different bases from appearing side by side. The code, the snapshot rebuild scripts and the
benchmark will be released together with the first full results; the release licence and location
are to be decided.

% References are inlined so the document compiles without external files.
\section*{References}
{\renewcommand{\section}[2]{}%
}

\end{document}